\documentclass[preprintnumbers,prd,twocolumn,showpacs,floatfix,nofootinbib,
superscriptaddress]{
revtex4}
 \pdfoutput=1
\usepackage{graphicx}
\usepackage{epsfig}
\usepackage{bm}
\usepackage{amssymb}
\usepackage{float}
\usepackage{amsmath}
\usepackage{dcolumn}
\usepackage{cancel}
\usepackage[colorlinks]{hyperref}
\usepackage[usenames,dvipsnames]{color}
\hypersetup{
     breaklinks=true,
    pdfstartview={FitH},    % fits the width of the page to the window
    colorlinks=true,       % false: boxed links; true: colored links
    linkcolor=blue,          % color of internal links
    citecolor=red,        % color of links to bibliography
    filecolor=magenta,      % color of file links
    urlcolor=blue,           % color of external links
    anchorcolor=green,      % Color for anchor text
    linktocpage=true
}

\def\doi{http://doi.org}
\def\arxiv{http://arxiv.org/abs}

\begin{document}

\title{Bayesian analysis of $f(T)$ gravity using $f\sigma_8$  data}

\author{Fotios K. Anagnostopoulos }\email{fotis-anagnostopoulos@hotmail.com}
\affiliation{Department of Physics, National \& Kapodistrian University of Athens, 
Zografou Campus GR 157 73, Athens, Greece}

\author{Spyros Basilakos}\email{svasil@academyofathens.gr}
\affiliation{Academy of Athens, Research Center for Astronomy and
Applied Mathematics, Soranou Efesiou 4, 11527, Athens, Greece}

\author{Emmanuel N. Saridakis}
\email{msaridak@phys.uoa.gr}
\affiliation{Department of Physics, National Technical University of Athens, Zografou
Campus GR 157 73, Athens, Greece}
\affiliation{Department of Astronomy, School of Physical Sciences, University of Science 
and Technology of China, Hefei 230026, P.R. China}
\affiliation{Chongqing University of Posts \& Telecommunications, Chongqing, 400065, P.R.
China}


%%%%%%%%%%%%%%
\begin{abstract}
We use observational data from Supernovae (SNIa) Pantheon sample, from direct Hubble 
constant measurements with cosmic chronometers (CC), from the 
Cosmic Microwave Background shift parameter $\text{CMB}_{\text{shift}}$,  and from 
redshift space distortion 
($f\sigma_8$) measurements, in order to constrain $f(T)$ gravity.  
We do not follow the common $\gamma$ parameterization within the semi-analytical 
approximation of the growth rate, in order to avoid model-dependent uncertainties. Up to 
our knowledge this is the first time that   $f(T)$ gravity is analyzed within a Bayesian 
framework, and with background and perturbation behaviour considered jointly. We show 
that all three examined  $f(T)$ models  are able to describe 
adequately the $f\sigma_8$ data. Furthermore, applying the Akaike, Bayesian  and 
Deviance Information Criteria, we conclude that all considered models   are 
statistically equivalent, however the most efficient candidate is the exponential model, 
which additionally presents  a small deviation from $\Lambda$CDM paradigm.
\end{abstract}

\maketitle

\section{Introduction}

The increasing collection of high accuracy data from cosmological observations, at both 
  background and perturbation levels, as well as the existing theoretical arguments, 
led to an enhanced interest in investigating the possibility that the fundamental 
gravitational theory is not general relativity but a modified theory which accepts the 
latter as a low-energy limit \cite{Capozziello:2011et,Nojiri:2010wj}.
Amongst the various  modified gravity  constructions one may have   
torsional  gravity (for a 
review see \cite{Cai:2015emx}),  which arises from an 
extension of the Teleparallel Equivalent of 
General Relativity (TEGR) \cite{ein28,Hayashi79,Pereira.book,Maluf:2013gaa}. Hence, 
 one can 
construct modifications such as   $f(T)$ gravity 
\cite{Cai:2015emx,Bengochea:2008gz,Linder:2010py,Chen:2010va,
Myrzakulov:2010tc,Zheng:2010am,Bamba:2010wb,Cai:2011tc,Li:2011rn,
Capozziello:2011hj, Wu:2011kh, Wei:2011aa, Amoros:2013nxa,
Otalora:2013dsa,Bamba:2013jqa,
Li:2013xea,Ong:2013qja,Nashed:2014lva,Darabi:2014dla, Haro:2014wha,
Guo:2015qbt, 
Bamba:2016gbu,Malekjani:2016mtm,Farrugia:2016qqe,
Qi:2017xzl,
Bahamonde:2017wwk,Karpathopoulos:2017arc,Abedi:2018lkr,DAgostino:2018ngy,Krssak:2018ywd,
Iosifidis:2018zwo,
Chakrabarti:2019bed, DavoodSadatian:2019pvq},    
$f(T,T_G)$ gravity 
\cite{Kofinas:2014owa,Kofinas:2014daa}, scalar-torsion theories 
\cite{Geng:2011aj,Hohmann:2018rwf}, etc.

Perhaps the most  crucial question in every  modified gravity is the determination of the 
involved  arbitrary function. Although some general features can be extracted through 
theoretical arguments, such as the existence of Noether symmetries, the absence of 
ghosts, 
the stability of perturbations, etc, the basic tool that one has is the confrontation 
with observations. In these lines, in the case of $f(T)$ gravity there has been a large 
amount of research towards this direction using solar system  data  
\cite{Iorio:2012cm,Iorio:2015rla,Farrugia:2016xcw}, gravitational waves data 
\cite{Cai:2018rzd,Nunes:2018evm,Nunes:2019bjq}, as well as cosmological ones
\cite{Wu:2010mn,Cardone:2012xq,Nesseris:2013jea,Capozziello:2015rda,Basilakos:2016xob,
Nunes:2016qyp,Nunes:2016plz,Nunes:2018xbm,Basilakos:2018arq,Xu:2018npu,Yang:2018euj}.
 
Up to now, the confrontation with cosmological data used mainly expansion data, namely 
data related to the background evolution, such that  Supernovae type Ia
data  (SNIa),  Cosmic Microwave Background (CMB) shift parameters,
Baryonic Acoustic Oscillations (BAO), and Hubble data observations. Large scale structure 
data were also applied, nevertheless they were used under the 
imposition of specific growth-index parameterizations  \cite{Nesseris:2013jea}. Hence, it 
would be interesting to investigate what would be the constraints on $f(T)$ gravity that 
arise from a Bayesian analysis   using $f\sigma_8$  data in a model-independent way, 
namely without assuming any form for the growth index.

In the present work we perform such a general analysis, in order to extract the 
constraints on $f(T)$ gravity from $f\sigma_8$  data. As we see, we obtain better 
constraints comparing to all other data sets apart from CMB shift parameter. 
Nevertheless, the interesting novel feature is that although the previous observational 
confrontation showed that the power-law, $f_{1}$CDM, model was the most well-fit one, the 
current analysis shows that the exponential, $f_{3}$CDM, model is the one that is 
preferred.

The plan of the work is the following:  In Section \ref{fTmodel} we present $f(T)$ 
gravity and we provide the cosmological equations at both background and perturbation 
levels. In Section \ref{DataMethodology} we present the various datasets and the 
methodology that we use. Then,  in Section \ref{results} we present the obtained 
results and the corresponding contour plots. Finally,   Section \ref{Conclusions} is 
devoted to the 
conclusions.

\section{$f(T)$ gravity and cosmology}
\label{fTmodel}

In this section we  review the cosmological equations in the framework of $f(T)$ gravity.
For its formulation one uses the vierbeins fields ${\mathbf{e}_A(x^\mu)}$ as dynamical 
variables, which at a manifold point  $x^\mu$ form an orthonormal
basis   ($\mathbf{e} _A\cdot \mathbf{e}_B=\eta_{AB}$ with $\eta_{AB}={\rm diag} 
(1,-1,-1,-1)$). In a coordinate basis they read  as
$\mathbf{e}_A=e^\mu_A\partial_\mu $ and the metric is given by
\begin{equation}  \label{metricrel}
g_{\mu\nu}(x)=\eta_{AB}\, e^A_\mu (x)\, e^B_\nu (x),
\end{equation}
with Greek  and Latin indices used for the coordinate and
tangent space respectively. Concerning the connection one introduces the 
 Weitzenb\"{o}ck one, namely
$\overset{\mathbf{w}}{\Gamma}^\lambda_{\nu\mu}\equiv e^\lambda_A\:
\partial_\mu
e^A_\nu$ \cite{Weitzenb23}, and thus the corresponding torsion tensor becomes
\begin{equation}
\label{torsten}
{T}^\lambda_{\:\mu\nu}\equiv\overset{\mathbf{w}}{\Gamma}^\lambda_{
\nu\mu}-%
\overset{\mathbf{w}}{\Gamma}^\lambda_{\mu\nu}
=e^\lambda_A\:(\partial_\mu
e^A_\nu-\partial_\nu e^A_\mu).
\end{equation}
The torsion tensor contains all the information of the gravitational field, and its  
contraction provides the torsion scalar 
\begin{equation}
\label{torsiscal}
T\equiv\frac{1}{4}
T^{\rho \mu \nu}
T_{\rho \mu \nu}
+\frac{1}{2}T^{\rho \mu \nu }T_{\nu \mu\rho }
-T_{\rho \mu }^{\ \ \rho }T_{\
\ \ \nu }^{\nu \mu },
\end{equation}
which forms the Lagrangian of teleparallel gravity (in similar lines to the fact that  
the Ricci scalar forms the Lagrangian of general relativity). 
Variation of the teleparallel action in terms of the vierbeins gives the same equations 
with general relativity, and thus the constructed theory was named teleparallel 
equivalent of general relativity (TEGR).

One can use TEGR as the starting point of gravitational modifications. The simplest
direction is to generalize $T$ to 
a function $T+f(T)$ in the action, namely
\cite{Cai:2015emx}
\begin{eqnarray}
\label{action0}
I = \frac{1}{16\pi G}\int d^4x e \left[T+f(T)+L_m\right],
\end{eqnarray}
with $e = \text{det}(e_{\mu}^A) = \sqrt{-g}$,  $G$ the gravitational
constant (we set the light speed to 1 for simplicity), and where we have also included 
the total matter Lagrangian $L_m$ for completeness.  
Varying the above action  we extract the field equations:
\begin{eqnarray}\label{equationsom}
&&e^{-1}\partial_{\mu}(ee_A^{\rho}S_{\rho}{}^{\mu\nu})[1+f_{T}]
 +
e_A^{\rho}S_{\rho}{}^{\mu\nu}\partial_{\mu}({T})f_{TT}\ \ \ \ \  \ \ \ \  \ \
\ \ \nonumber\\
&& \ \ \ \
-[1+f_{T}]e_{A}^{\lambda}T^{\rho}{}_{\mu\lambda}S_{\rho}{}^{\nu\mu}+\frac{1}{4} e_ {A
} ^ {
\nu
}[T+f({T})] \nonumber \\
&&
\ \ \ \ \,
= 4\pi Ge_{A}^{\rho}\overset {\mathbf{em}}T_{\rho}{}^{\nu},
\end{eqnarray}
where we have defined $f_{T}\equiv\partial f/\partial T$, $f_{TT}\equiv\partial^{2} 
f/\partial T^{2}$,
and moreover $\overset{\mathbf{em}}{T}_{\rho}{}^{\nu}$  stands for  the total matter 
(i.e. 
baryonic and dark matter and radiation) 
energy-momentum 
tensor. Additionally,  we have introduced the ``super-potential''
$
S_\rho^{\:\:\:\mu\nu}\equiv\frac{1}{2}\Big(K^{\mu\nu}_{\:\:\:\:\rho}
+\delta^\mu_\rho
\:T^{\alpha\nu}_{\:\:\:\:\alpha}-\delta^\nu_\rho\:
T^{\alpha\mu}_{\:\:\:\:\alpha}\Big)$, where 
$K^{\mu\nu}_{\:\:\:\:\rho}\equiv-\frac{1}{2}\Big(T^{\mu\nu}_{
\:\:\:\:\rho}
-T^{\nu\mu}_{\:\:\:\:\rho}-T_{\rho}^{\:\:\:\:\mu\nu}\Big)$ is  the contorsion tensor.

\subsection{Background behavior}

In order to proceed to the cosmological application of $f(T)$ gravity we impose 
the homogeneous and isotropic  flat Friedmann-Robertson-Walker (FRW) geometry 
\begin{equation}
ds^2= dt^2-a^2(t)\,  \delta_{ij} dx^i dx^j,
\end{equation}
which corresponds to the vierbein choice 
$e_{\mu}^A={\rm
diag}(1,a,a,a)$,
with $a(t)$ the scale factor. Inserting this choice  
into (\ref{equationsom}) we extract the Friedmann equations for $f(T)$ cosmology as
\begin{eqnarray}\label{Fr11}
&&H^2= \frac{8\pi G}{3}(\rho_m+\rho_r)
-\frac{f}{6}+\frac{Tf_T}{3}\\\label{Fr22}
&&\dot{H}=-\frac{4\pi G(\rho_m+P_m+\rho_r+P_r)}{1+f_{T}+2Tf_{TT}},
\end{eqnarray}
where 
$H\equiv\dot{a}/a$  is the Hubble function and dots denote
derivatives with respect to $t$. Moreover, in the 
above equations $\rho_m$, $\rho_r$ and   $P_m$, $P_r$ are
the energy densities and pressures of the matter and radiation sectors respectively, 
which are considered to constitute the total matter energy-momentum tensor. Finally, 
note that  in FRW geometry the torsion scalar  (\ref{torsiscal}) becomes  $T=-6H^2$, and 
such an interchanging relation between $T$ and $H^2$ proves to be very helpful.

Observing the form of the first Friedmann equation (\ref{Fr11}) we deduce that we can 
define an effective dark energy 
sector with   energy density and pressure respectively given by
\begin{eqnarray}
&&\rho_{DE}\equiv\frac{3}{8\pi
G}\left[-\frac{f}{6}+\frac{Tf_T}{3}\right], \label{rhoDDE}\\
\label{pDE}
&&P_{DE}\equiv\frac{1}{16\pi G}\left[\frac{f-f_{T} T
+2T^2f_{TT}}{1+f_{T}+2Tf_{TT}}\right],
\end{eqnarray}
and thus its equation-of-state parameter becomes
\begin{eqnarray}
\label{wefftotf}
 w\equiv\frac{P_{DE}}{\rho_{DE}}
=-\frac{f/T-f_{T}+2Tf_{TT}}{\left[1+f_{T}+2Tf_{TT}\right]\left[f/T-2f_{T}
\right] }.
\end{eqnarray}
We mention that the cosmological equations close   by   considering the conservation  
equations  of matter 
and radiation sectors:
\begin{eqnarray}
\label{mattradevol}
 \dot{\rho}_m+3H(\rho_m+P_m)=0\\
  \dot{\rho}_r+3H(\rho_r+P_r)=0,
  \label{evoleqr}
\end{eqnarray}
which according to  (\ref{Fr11}), (\ref{Fr22})
then imply the conservation of the effective dark-energy sector too, namely  
\begin{eqnarray}
  \dot{\rho}_{DE}+3H(\rho_{DE}+P_{DE})=0.
  \label{evoleqr2}
\end{eqnarray}

In order to  elaborate the modified Friedmann equations, following 
\cite{Nesseris:2013jea,Basilakos:2018arq}  
we introduce
\begin{eqnarray}
\label{Edefinition}
E^{2}(z)\equiv\frac{H^2(z)} {H^2_{0}}=\frac{T(z)}{T_{0}},
\end{eqnarray}
where $T_0\equiv-6H_{0}^{2}$, with $H_0$ the present value of the Hubble function (from 
now on the subscript ``0'' denotes the value of a quantity at present). Additionally, as 
the independent variable
we use the redshift $z=\frac{a_0}{a}-1$, with $a_0$   the current scale 
factor set to one for simplicity. As usual, we consider the matter sector to be dust, 
namely $w_m\equiv P_m/\rho_m=0$, and thus    (\ref{mattradevol}) implies that 
$\rho_{m}=\rho_{m0}(1+z)^{3}$, and similarly imposing for the radiation sector
$w_r\equiv P_r/\rho_r=1/3$ from  (\ref{evoleqr}) we obtain   
$\rho_{r}=\rho_{r0}(1+z)^{4}$.
Hence, the Friedmann equation (\ref{Fr11}) can be expressed as
\begin{eqnarray}
\label{Fr11EZrel}
E^2(z,{\bf r})=\Omega_{m0}(1+z)^3+\Omega_{r0}(1+z)^4+\Omega_{F0} y(z,{\bf r}),\
\end{eqnarray}
where  
\begin{equation}
\label{distortionpar}
 y(z,{\bf r})=\frac{1}{T_0\Omega_{F0}}\left[f-2Tf_T\right].
\end{equation}
In these expressions we have introduced the density parameters 
$\Omega_{i}=\frac{8\pi G \rho_{i}}{3H^2}$, with $\Omega_{m0}$, $\Omega_{r0}$ their 
present values, and we have defined
\begin{equation}
\label{OmF00}
\Omega_{F0}=1-\Omega_{m0}-\Omega_{r0} \;.
\end{equation}
Hence,   the effect of $f(T)$ gravity at the background level 
  is quantified by the function 
 $y(z,{\bf r})$, normalized to
unity at   present time. This quantity depends on $\Omega_{m0}$ and $\Omega_{r0}$, as 
well as  
on the free
  parameters $r_1,r_2,...$, assembled to the vector $\bf r$ that a specific $f(T)$ 
model includes (the exact elements of the $\bf r$ vector are defined later on). 
Finally, as expected, in the limit of $\Lambda$CDM 
cosmology, i.e. for  $f(T)=const.$, the function 
$y(z,{\bf r})$ is just a constant.

\subsection{Linear matter perturbations}

In any cosmological model that does not include interactions in the dark sector, 
at  sub-horizon scales and through matter epoch, 
the basic equation that determines the evolution of the matter
perturbations in the linear regime is 
\cite{BasNes13,Gannouji:2008wt,Lue:2004rj,Linder:2005in,Stabenau:2006td,Uzan:2006mf,
Tsujikawa:2007tg}
\begin{equation}
\label{odedelta}
\ddot{\delta}_{m}+ 2H\dot{\delta}_{m}=4 \pi G_{\rm eff} \rho_{m} \delta_{m},
\end{equation}
where $\delta_{m}\equiv\delta\rho_{m}/\rho_m$ is the matter overdensity. In the above 
equation one introduces the effective Newton's   constant
  $G_{\rm eff}(a)=G_{N} Q(a)$, with $G_{N}$ 
the  gravitational constant appearing in the action of the theory, which reflects the 
information of the gravitational modification. In general  $G_{\rm eff}(a)$ is varying, 
and the specific form of $Q(a)$ is determined by the underlying gravitational theory.
For general-relativity  we have $G_{\rm eff}(a)=G_{N}$ (i.e. $Q(a)=1$) 
and thus (\ref{odedelta}) provides the usual evolution
equation for matter over-density \cite{Peeb93}.

From the above discussion it becomes obvious that we can apply this general perturbation 
treatment in the case of $f(T)$ cosmology, as long as we know the 
form of $G_{\rm eff}(a)$, or 
equivalently $Q(a)$, of $f(T)$ gravity. It is relatively easy to show that for $f(T)$ 
gravity \cite{Zheng:2010am,Nesseris:2013jea}
\begin{eqnarray}
\label{Geff}
Q(a)=\frac{G_{\rm eff}(a)}{G_{N}}=\frac{1}{1+f_{T}},
\end{eqnarray}
 a relation that arises from the complete perturbation analysis \cite{Chen:2010va}.
Note that this expression is significantly simpler than the corresponding one of 
$f(R)$ gravity, since the latter includes a scale dependence.

Let us make a comment here, on the usual handling of perturbation growth in the 
literature. 
In order to confront the theoretical calculations with observations it is  
common practice to 
introduce the clustering growth rate  as 
\cite{Peeb93}
\begin{equation}
\label{growthrate}
F(a)=\frac{d\ln \delta_{m}}{d\ln a}\simeq \Omega^{\gamma}_{m}(a),
\end{equation}
where $\gamma$ is the growth index. In the case of  dark energy scenarios in the 
framework of general relativity, with constant  equation-of-state parameter $w$,  the 
growth index is
well approximated by $\gamma \simeq \frac{3(w-1)}{6w-5}$
\cite{Silveira:1994yq,Wang:1998gt,Lue:2004rj,Linder:2004ng,Linder:2007hg,Nesseris:2007pa},
which for $\Lambda$CDM cosmology ($w=-1$) reduces to $\gamma \approx 6/11$.
 Inserting  (\ref{growthrate}) into  Eq. (\ref{odedelta}) we find
\begin{equation}
\label{fzz222}
a\frac{dF(a)}{da}+F(a)^{2}+X(a)F(a)
= \frac{3}{2}\Omega_{m}(a)Q(a) \;,
\end{equation}
with
\begin{equation}
\label{xxarel}
X(a)=\frac{1}{2}-\frac{3}{2}w(a)
\left[ 1-\Omega_{m}(a)\right] ,
\end{equation}
where we have used the relations   \cite{BasNes13,Nesseris:2013jea}
\begin{equation}
\label{eos222}
w(a)=\frac{-1-\frac{2}{3}a\frac{{d\rm lnE}}{da}}
{1-\Omega_{m}(a)} \;,
\end{equation}
\begin{equation}
\label{ddomm}
\Omega_{m}(a)=\frac{\Omega_{m0}a^{-3}}{E^{2}(a)} \,,
\end{equation}
and thus
\begin{equation}
\label{domm}
\frac{d\Omega_{m}(a)}{da}=
\frac{3}{a}w(a)\Omega_{m}(a)\left[1-\Omega_{m}(a)\right]\;.
\end{equation}
We would like to mention that the above semi-analytical approximation of 
the 
growth rate, although convenient and useful at specific investigations, seems to reduce 
the numerical burden of the analysis and also to serve as a null diagnostic for the 
nature 
of dark energy. However, for different models than the concordance $\Lambda$CDM one, the 
approximation error increases as a function of the model parameters. This property could 
possibly flaw the extracted parameter values and the subsequent model selection. 
Furthermore, one needs to add at least one extra free parameter to the likelihood 
analysis. In summary, for the above   reasons, in the following we prefer    not to 
use this approximation and use the $f\sigma_8$  data in a completely model-independent, 
Bayesian way. In this way, namely using the 
full numerical solution of \eqref{odedelta} instead of  the growth index ($\gamma(z)$) 
semi-analytical approximation,
we have the  advantage of  a reduced numerical error, as well as   the 
independence from a certain gamma parameterization.

\subsection{Specific $f(T)$ models}
\label{fTmodels}

We close this section by presenting three specific viable $f(T)$ models with two 
parameters, one of which is independent, i.e models that are efficient in 
successfully passing the confrontation with observations 
\cite{Nesseris:2013jea,Basilakos:2018arq}. Furthermore, we  quantify their deviation 
from $\Lambda$CDM paradigm in a unified way, through  the
function $y(z,{\bf r})$ of (\ref{distortionpar}) and a   distortion
parameter $b$.  Hence, the elements of the $\bf r$ vector, namely the 
parameters $r_1$and $r_2$, for all the following 
models are $\Omega_{m0}$ and $b$.

\begin{enumerate}
\item The power-law model \cite{Bengochea:2008gz}
(hereafter $f_{1}$CDM model), in which 
\begin{equation}
f(T)=\alpha (-T)^{b}.
\label{powermod}
\end{equation} 
Inserting it
into  (\ref{Fr11}) at present time we
find
\begin{eqnarray}
\alpha=(6H_0^2)^{1-b}\frac{\Omega_{F0}}{2b-1},
\end{eqnarray}
and hence the only free parameter is $b$. Additionally,
 (\ref{distortionpar})
leads to
\begin{equation}
\label{yLL}
y(z,b)=E^{2b}(z,b) \;.
\end{equation}
 Thus, for $b=0$ the model  $f_{1}$CDM  
reduces to $\Lambda$CDM cosmology, i.e. 
$T+f(T)=T-2\Lambda$, with $\Lambda=3\Omega_{F0}H_{0}^{2}$ and
$\Omega_{F0}=\Omega_{\Lambda 0}$.

\item The square-root exponential model (hereafter $f_{2}$CDM) \cite{Linder:2010py}
\begin{eqnarray}
f(T)=\alpha T_{0}(1-e^{-p\sqrt{T/T_{0}}}).
\label{Lindermod}
\end{eqnarray}
 In this case  Eq.
(\ref{Fr11}) at present gives
\begin{eqnarray}
\alpha=\frac{\Omega_{F0}}{1-(1+p)e^{-p}},
\end{eqnarray}
while (\ref{distortionpar}) leads to
\begin{equation}
\label{yLLf2}
y(z,p)=\frac{1-(1+pE)e^{-pE}}{1-(1+p)e^{-p}}.
\end{equation}
   $f_{2}$CDM model reduces
to  $\Lambda$CDM paradigm  for $p \rightarrow +\infty$, and thus we can 
replace $p$ by $p=1/b$, acquiring
\begin{equation}
\label{yf2}
y(z,b)=\frac{1-(1+\frac{E}{b})e^{-E/b}}{1-(1+\frac{1}{b})e^{-1/b}},
\end{equation}
which tends to 1 for
$b \rightarrow 0^{+}$.

\item  The   exponential model (hereafter $f_{3}$CDM) \cite{Nesseris:2013jea}:
\begin{eqnarray}
f(T)=\alpha T_{0}(1-e^{-pT/T_{0}}).
\label{f3cdmmodel}
\end{eqnarray}
In  this case
\begin{eqnarray}
\alpha=\frac{\Omega_{F0}}{1-(1+2p)e^{-p}},
\end{eqnarray}
and 
\begin{equation}
\label{modfcdm3}
y(z,p)=\frac{1-(1+2pE^{2})e^{-pE^{2}}}{1-(1+2p)e^{-p}}.
\end{equation}
Finally, we may re-write  these expressions using
 $p=1/b$, obtaining
\begin{equation}
\label{modfcdm3b}
y(z,b)=\frac{1-(1+\frac{2E^{2}}{b})e^{-E^{2}/b}}{1-(1+\frac{2}{b})e^{-1/b}},
\end{equation}
which implies that for $b \rightarrow 0^{+}$ the $f_{3}$CDM model reduces to  
$\Lambda$CDM one.

\end{enumerate}

\section{Data and Methodology}
\label{DataMethodology}

In this section we first present the various data sets and subsequently we describe the 
statistical 
methods that we employ. 
In particular, we use $f\sigma_{8}$ data, data from direct measurements of the
Hubble parameter, and data from standard candles (SNIa).
 As a next step, we assess the quality of the fit with the aid of 
various information criteria.  In what follows, we present explicitly the aforementioned 
steps.

\subsection{Cosmological probes}

\subsubsection{f$\sigma$8 data}

An almost model-independent cosmological probe, namely the $f \sigma_{8}$ product, arises 
from the analysis of redshift-space distortions \cite{Song:2008qt}. In the aforementioned 
product, $f(z)$ is the growth rate of clustering and $\sigma_8$ is the effective variance 
of the density function within spheres of radius $8 \ h^{-1} Mpc$, where linear 
perturbations is a good approximation. There is a large number of data points available 
in 
the literature. Hence, a usual problem that appears is that the degree 
of overlap between surveys is in 
general unknown, thus there are unknown correlations between the data points, which in 
turn makes the standard joint likelihood analysis unsuitable. For the above reasons, we 
choose to use a compilation of $f\sigma_8$ data that has been explicitly checked in terms 
of its robustness using information theoretical methods (see \cite{Sagredo:2018ahx} and 
in particular their Table I, with the corresponding references). The relevant 
chi-square function reads
\begin{equation}
    \chi^2_{f\sigma 8} = 
\sum_{i=1}^{22}\left(\frac{f\sigma_{8,obs,i}-f\sigma_{8}(a_{i},\phi^{\nu+1})_{theor}}
{\sigma_i}\right)^2,
\end{equation}
where $f\sigma_{8}(a_{i},\phi^{\nu+1})_{theor}= 
\sigma_{8}\delta'(a_{i},\phi^{\nu})/\delta(1,\phi^{\nu})a_{i}$ and a prime 
denotes derivative of the scale factor $a$.
The quantity $\sigma_{8}$ is a free parameter. The statistical vector $\phi^{\nu}$  
contains the free parameters of the statistical model under consideration, which 
are the elements of the $\bf r$ vector plus the Hubble constant $H_0$, the $\sigma_8$,
and the hyper-parameters that are described latter in the text. The values 
$\delta'(a_{i})$, $\delta(1)$ are calculated by the numerical solution of Eq. 
\eqref{odedelta} for a given set of cosmological parameters.

\subsubsection{Direct measurements of the Hubble expansion}
 	
          From the latest $H(z)$ data set compilation available, Ref. \cite{YuRatra2018}, 
we use  only  data obtained from  cosmic chronometers (CC).  These are massive 
galaxies evolving ``slowly" at certain intervals of the cosmic time. By using their 
differential age, one can measure the Hubble rate directly (see e.g. Ref. 
\cite{Moresco:2018xdr} and references therein). A striking advantage of the differential 
age of passive evolving galaxies is that the resulting measurement of the Hubble rate 
comes without any assumptions for the underlying cosmology. Our study incorporates $N=31$ 
measurements of the Hubble expansion in the redshift range $0.07 \lesssim z \lesssim 2.0 
$.
 
Here, the corresponding $\chi^2_{H}$ function reads
          \begin{equation}
          \chi^{2}_{H}\left(\phi^{\nu}\right)={\bf \cal H}\,
          {\bf C}_{H,\text{cov}}^{-1}\,{\bf \cal H}^{T}\,,
          \end{equation}
          where ${\bf \cal H }=\{H_{1}-H_{0}E(z_{1},\phi^{\nu})\,,\,...\,,\,
          H_{N}-H_{0}E(z_{N},\phi^{\nu})\}$ and $H_{i}$ are the observed Hubble rates at 
 redshift $z_{i}$ ($i=1,...,N$).

 \begin{table*}[!]
\tabcolsep 4.pt
\vspace{1mm}
\begin{tabular}{ccccccccc} \hline \hline
Model & $\Omega_{m0}$ & $h$ & $b $  & $\sigma_{8}$ & $\mathcal{M}$ &
$\Omega_{b}h^2$ & $\chi_{\text{min}}^{2} $&  $\chi_{\text{min}}/dof$  \vspace{0.05cm}\\ 
\hline
%----------------------------------------0.02227_{-0.00016}^{+0.00016} 
\hline
  \vspace{0.01cm}\\
 \multicolumn{8}{c}{\emph{$H(z)$ + \text{SNIa} + \text{$f\sigma_{8}$}}}\\ \\
  $f_1$ & $0.291_{-0.029}^{+0.034}   $ & $0.6921_{-0.0181}^{+0.0185}  $ & 
$0.021_{-0.249}^{+0.183}$ & $ 0.778_{-0.063}^{+0.080} $ &$ -19.378\pm{0.054}$& - & $ 
66.968 
$ 
& 0.761\\ %%NAI
  
 $f_2$ & $0.282_{-0.029}^{+0.024}$ & $0.693 \pm 0.018$ & $0.180_{-0.133}^{+0.176}$ & 
$0.789_{-0.041}^{+0.051}$ & $-19.372_{-0.053}^{+0.054} $ & - & $69.000$ & 0.784\\ %% OK
 
 $f_3$  & $ 0.290 \pm 0.020  $ & $0.6928_{-0.020}^{+0.018}  $ & $  
0.097_{-0.070}^{+0.074} $& $0.775 \pm 0.035$ &$ -19.374\pm{0.053} $ & - & $ 67.767 $ &   
0.770\\ %% OK

$\Lambda$CDM & $0.293_{-0.019}^{+0.020}$ & $0.6929 _{-0.0180}^{+0.0184} $& - & 
$0.769_{-0.033}^{+0.033}$ & $-19.376 
\pm 0.053$ & - & 67.019 & 0.753  \\  
 \vspace{0.01cm}\\
 \multicolumn{8}{c}{\emph{$H(z)$ + \text{SNIa} + \text{$f\sigma_{8}$} + 
$\text{CMB}_{\text{shift}}$}}\\ \\
 %all here are random numbers
 $f_1$ & $0.302 \pm 0.0110 $ & $0.6860\pm 0.0114    $ & 
$-0.063_{-0.087}^{+0.076}$ & $0.753_{-0.030}^{+0.032} $ &$ -19.399 \pm 0.027$ &$0.0223\pm 
0.0002 $ & $  67.707 
$ 
& 0.752\\ %
  
 $f_2$ & $0.310 \pm 0.008$ & $0.6780_{-0.0063}^{+0.0064}$ & $0.095_{-0.061}^{+0.073}$ & 
$0.758_{-0.030}^{+0.032}$ & $-19.419_{-0.016}^{+0.017} $ & $0.0223 \pm 0.0002 $ & $69.306$ 
&  0.797\\ %% OK
 
 $f_3$  & $ 0.309 \pm 0.008 $ & $ 0.6770_{-0.0062}^{+0.0064}  $ & $  
0.081_{-0.051}^{+0.058} $& $0.793 \pm 0.035$ &$ -19.418 \pm 0.016  $& $0.0223 \pm 0.0002 
$ 
& $ 68.967 $ &   
0.793\\ 

$\Lambda$CDM & $0.309 \pm 0.008 $ & $0.6777_{-0.0058}^{+0.0060} $& - & 
$0.757_{-0.030}^{+0.032}$ & $-19.417 \pm 0.016$ & $0.0223_{-0.0002}^{+0.0001}$ & 68.110 & 
0.748  \\ 
  \\ 
\hline\hline
\end{tabular}
\caption[]{Observational constraints and the
corresponding $\chi^{2}_{\rm min}$ for the power-law $f_1$CDM model  
(\ref{powermod}), for the  square-root exponential model  $f_2$CDM model 
(\ref{Lindermod}), and for the  exponential model  $f_3$CDM model
(\ref{f3cdmmodel}), using CC/Pantheon/f$\sigma_8$ and 
CC/Pantheon/f$\sigma_8$/$\text{CMB}_{\text{shift}}$ datasets.  For 
direct comparison we additionally include the concordance $\Lambda$CDM scenario.}
\label{tab:Results1}
\end{table*}

\subsubsection{Standard Candles}

A ``standard" or ``standarizable" candle is a luminous extra-galactic 
astrophysical object with observable features that are independent of the cosmic time. 
The 
most studied standard candles are, arguably, Supernovae Type Ia (SNIa). We include in our 
analysis the most recent SNIa set data available, namely the binned Pantheon sample of 
Scolnic et. al. \cite{Scolnic:2017caz}. As discussed in the latter, the full dataset
is very well approximated with the binned dataset of N = 40 data 
points in the redshift range $0.01 \lesssim z \lesssim 1.6$.
          The chi-square function 
of the SNIa data is given by
          \begin{equation}
          \chi^{2}_{SN Ia}\left(\phi^{\nu}_{\text{SNIa}}\right)={\bf \mu}_{\text{SNIa}}\,
          {\bf C}_{\text{SNIa},\text{cov}}^{-1}\,{\bf \mu}_{\text{SNIa}}^{T}\,,
          \end{equation}
          where
          ${\bf 
\mu}_{\text{\text{SNIa}}}=\{\mu_{1}-\mu_{\text{th}}(z_{1},\phi^{\nu})\,,\,...\,,\,
  \mu_{N}-\mu_{\text{th}}(z_{N},\phi^{\nu})\}$. The distance modulus is given as
  $\mu_{i} = \mu_{B,i}-\mathcal{M}$, where $\mu_{B,i}$ is the apparent magnitude 
at maximum in the rest frame for 
          redshift $z_{i}$, while the quantity $\mathcal{M}$ is a hyper-parameter (see  
\cite{Scolnic:2017caz} and references therein), quantifying uncertainties from various 
origins (astrophysical, data-reduction pipeline, etc). Furthermore, the theoretical
 form of the distance modulus reads
  \begin{equation}
  \mu_{\text{th}} = 5\log\left(\frac{d_{L}(z)}{\text{Mpc}}\right) + 25\,,
 \end{equation}
where
 \begin{equation}
d_L(z) = c(1+z)\int_{0}^{z}\frac{dx}{H(x,\phi^{\nu})}
\end{equation}
is the luminosity distance, for spatially flat FRWL geometry. It is apparent that 
$\mathcal{M}$ and the normalized Hubble constant $h$ are intrinsically degenerate in
   the context of the Pantheon data set, and therefore we are not in position to obtain 
any 
physical information regarding    $H_{0}$ from SNIa data alone.

\subsubsection{CMB shift data}
 The observations of temperature anisotropies in the CMB provide a valuable independent 
test for
the reality of dark energy at the recombination epoch $z \sim 1090$. The photons were 
coupled to
baryons and electrons before that red shift and decoupled right after. Due to the fact 
that in the
Boltzmann and Einstein equations all the components of the universe are coupled, in
order to extract information from the full spectrum, demanding numerical simulations are 
needed.
A convenient and efficient way to summarize information from the CMB data, without using 
the full
spectrum, is by employing the so-called CMB shift parameters or distance priors
\cite{HuSugiyama1996}. The main idea behind this approach is the simple fact that the 
impact of
the underlying cosmology is much more severe at certain features of the power spectrum 
such as the
position of a peak, as opposed to others, e.g. the shape of the curve in a slow-changing 
regime.

Following \cite{HuSugiyama1996}, we define
\begin{equation}
l_a = \pi \frac{r(z_{*})}{r_s(z_{*})}
\end{equation}
\begin{equation}
\mathcal{R} = \sqrt{\Omega_{m0}}H_{0}D_{A}(z_{*})c^{-1}\,.
\end{equation}
The quantity $d_{A}$ is the standard angular diameter distance equal to $d_A = 
d_L(1+z)^{-1}$ and
$r_s$ is the co-moving sound horizon defined as
\begin{equation}
r_{s}=\int_{0}^{t}\frac{c_s(t')dt'}{a(t')}=\frac{c}{H_{0}} 
\int_{0}^{a}\frac{c_s(a')da'}{E(a')a'^2},
\end{equation}
where the sound velocity is $c_s(a)=1/\sqrt{3(1+R_{b}a)}$ with
$R_{b}=31500\Omega_{b0}h^2(T_{\text{CMB}}/2.7K)^{-4}$ and $T_{\text{CMB}}=2.7255K$.
In order to obtain the redshift of the recombination epoch $z_{*}$ we use the following 
fitting formula
of \cite{HuSugiyama1996}
\begin{equation}
z_{*}=1048\left[1\!+\!0.00124(\Omega_{b0}h^2)^{-0.738}\right]\times
\left[1\!+\!g_1\left(\Omega_{m0} h^2\right)^{g_2}\right],
\end{equation}
where the quantities $g_1,g_2$ are defined as
\begin{equation}
g_1 = 
\frac{0.0783(\Omega_{b0}h^2)^{-0.238}}{1+39.5\left(\Omega_{b0}h^2\right)^{0.763}}\,\,\,
,
\,\,\,\,\, g_2=\frac{0.560}{1+21.1\left(\Omega_{b0}h^2\right)^{1.81}}\,.
\end{equation}
and $\Omega_{b0}$ is the mormalized baryon energy density today.

The relevant $\chi^2$ expression is
\begin{equation}
\chi^2_{\text{CMB}}=\left(\Delta l_a, \Delta \mathcal{R},
\Delta \Omega_{*} \right)\,C_{\text{cov}}^{-1}\,\left(\Delta l_a, \Delta \mathcal{R},
\Delta \Omega_{*} \right)^{T}\,,
\end{equation}
where $\Delta l_a = l_a - 301.77$, $\Delta \mathcal{R} = \mathcal{R} - 1.7482$,
$\Delta \Omega_{*} = \Omega_{b0} - 0.02226$. The corresponding uncertainties are
$\sigma_{l}=0.090$, $\sigma_{\mathcal{R}} = 0.0048$, $\sigma_{\Omega_{*}}=0.00016$, while
the covariance matrix is $C_{ij} = \sigma_{ij}c_{ij}$, where $\sigma_{ij}$ is
the uncertainty and $c_{ij}$ the elements of the normalized covariance matrix
taken from \cite{CMBshift_last}. We mention here that the aforementioned data are 
taken from the Planck 2015 results \cite{CMBshift_last}.

\begin{figure*}[!]
\centering\includegraphics[width=0.9\textwidth]{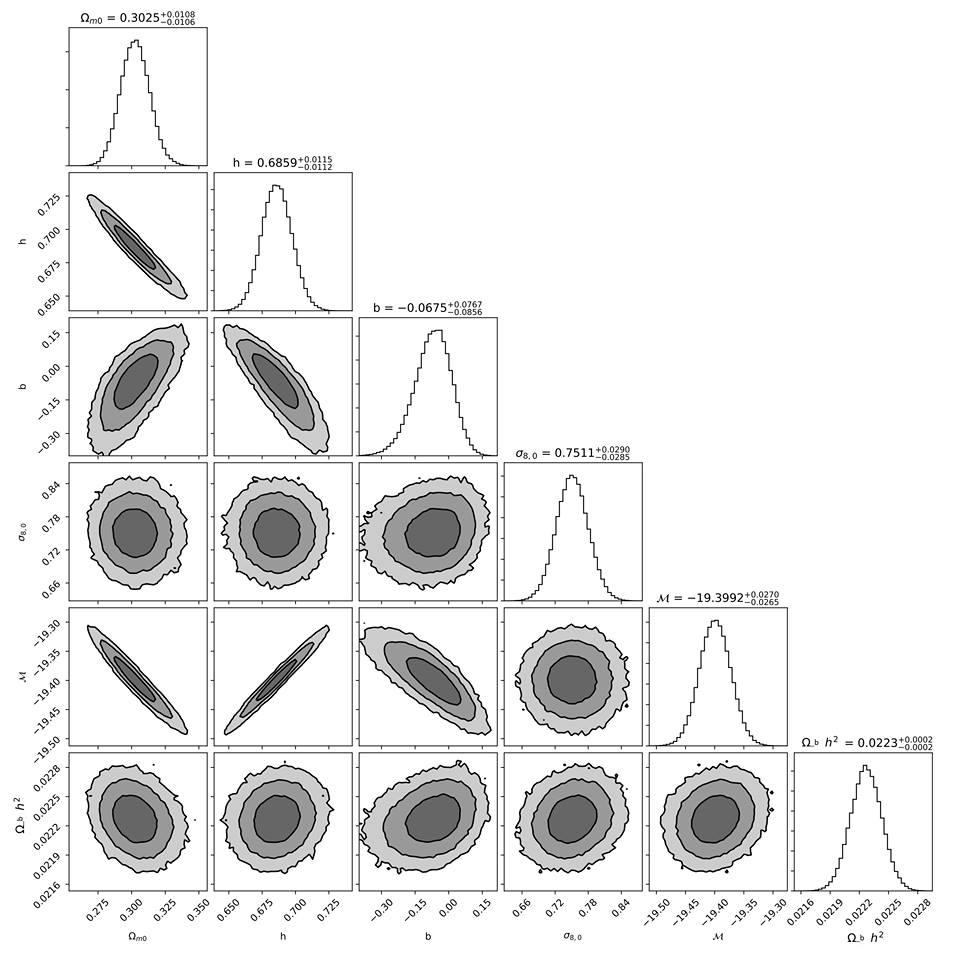}
\caption{{\it{The $1\sigma$, $2\sigma$ and $3\sigma$ iso-likelihood contours for  
the power-law $f_1$CDM model  
(\ref{powermod}), for all possible 2D subsets of the parameter space
$(\Omega_{m0},b,h,\sigma_8,\mathcal{M})$. Additionally, we provide the mean values
of the parameters within   the $1\sigma$ area of the MCMC chain. We have used joint 
analysis of CC/Pantheon/f$\sigma_8$/$\text{CMB}_{\text{shift}}$ data.}}}
\label{f_1_all}
\end{figure*}

  \begin{figure*}[!]
\centering\includegraphics[width=0.9\textwidth]{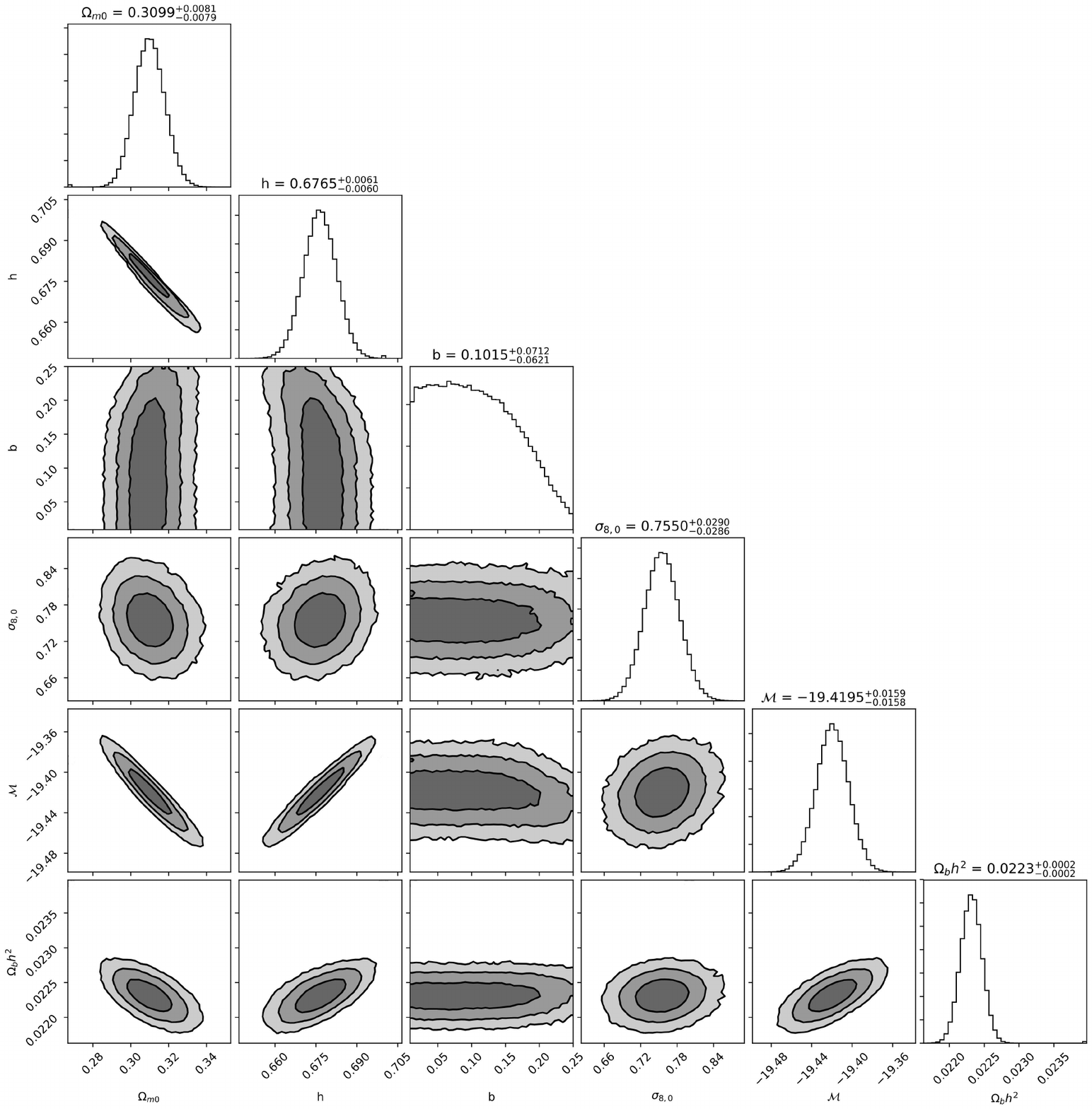}
\caption{{\it{The $1\sigma$, $2\sigma$ and $3\sigma$ iso-likelihood contours     
 for the  square-root exponential model  $f_2$CDM model 
(\ref{Lindermod}), for all possible 2D subsets of the parameter space
$(\Omega_{m0},b,h,\sigma_8,\mathcal{M})$. Additionally, we provide the mean values
of the parameters within   the $1\sigma$ area of the MCMC chain. We have used joint 
analysis of  CC/Pantheon/f$\sigma_8$/$\text{CMB}_{\text{shift}}$ data.}}}
\label{f_2_all}
\end{figure*}

\begin{figure*}[ht]
\centering\includegraphics[width=0.9\textwidth]{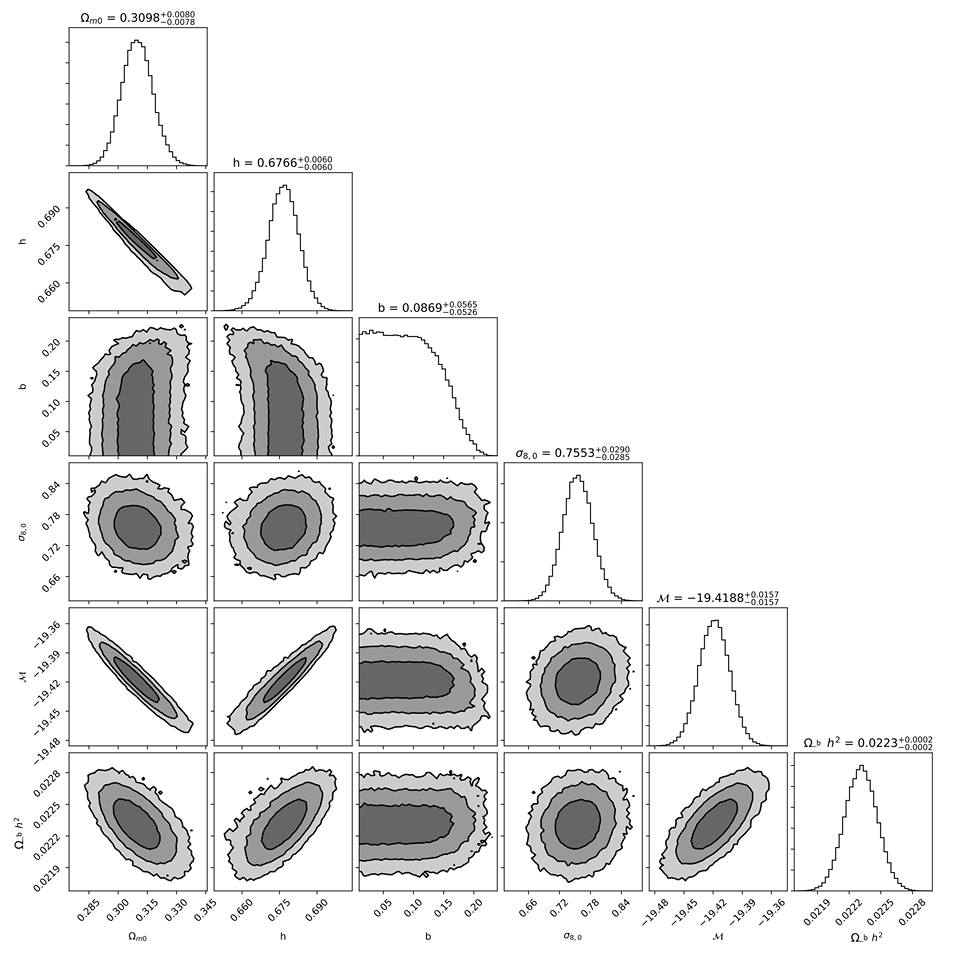}
\caption{{\it{The $1\sigma$, $2\sigma$ and $3\sigma$ iso-likelihood contours     
 for the  exponential model  $f_3$CDM model
(\ref{f3cdmmodel}), for all possible 2D subsets of the parameter space
$(\Omega_{m0},b,h,\sigma_{8,0},\mathcal{M})$. Additionally, we provide the mean values
of the parameters within   the $1\sigma$ area of the MCMC chain. We have used joint 
analysis of  CC/Pantheon/f$\sigma_8$/$\text{CMB}_{\text{shift}}$ data. }}}
\label{f_3_all}
\end{figure*}

\subsubsection{Joint analysis}

For obtaining the joint constraints on the cosmological parameters from  $P$ cosmological 
probes,
        we define the total likelihood function as follows
        \begin{equation}
        \mathcal{L}_{\text{tot}}(\phi^{\psi}) = \prod_{p=1}^{P} \exp(-\chi^2_{p})\,.
        \end{equation}
        Moreover, the corresponding  $\chi^2_{\text{tot}}$ expression is given by
        \begin{equation}
        \chi_{\text{tot}}^2 = \sum_{p=1}^{P}\chi^2_{P}\,.
        \end{equation}
        The statistical vector has dimension $k$, that is $\nu$
        parameters of the model at hand plus the number $\nu_{\text{hyp}}$ of 
hyper-parameters from the data sets used, resulting to $k = \nu + 
\nu_{\text{hyp}}$. Finally, the vector containing the free parameters that we 
constrain is $\phi^{\mu} = \{\Omega_{m0},h,b,\sigma_{8,0},\mathcal{M} \}$ for the 
$CC+SNIa+f\sigma_{8}$ dataset and  $\phi^{\mu} = 
\{\Omega_{m0},h,b,\sigma_{8,0},\mathcal{M}, \Omega_{b}h^2\}$ for the 
$CC+SNIa+f\sigma_{8}+CMB_{shift}$ dataset. However, there is no difference between the 
intrinsic hyper-parameters 
of a given data set and the free parameters of a cosmological model from the statistical 
perspective. Regarding the problem of likelihood maximization, we use an affine-invariant 
Markov Chain Monte Carlo sampler \cite{AffineInvMCMC}, as it is implemented within the 
open-source Python package emcee \cite{emcee}. We used 900 chains (walkers) and 3000 
steps (states). The convergence of the MCMC algorithm is checked with auto-correlation 
time considerations.

Finally, let us make the following comment. It is well known that the growth rate 
data may depend on the fiducial cosmological model utilized by various teams to convert
redshifts to distances. This is the so called Alcock-Paczynski effect.
In order to correct the data one has to rescale
the $f\sigma_{8}$ measurements and uncertainties by the ratios of $H(z) D_{A}(z)$ of the 
cosmology used, to that of the reference one \cite{Nesseris:2017vor,Macaulay:2013swa}.   
However, as found in \cite{Macaulay:2013swa} (see their Fig. 1),
the above correction is very small. Indeed, applying this correction in the case of the 
power-law $f_1$CDM  model of the present work we verify that the constraints are in 
agreement (within 1$\sigma$) with those based on the original published growth data.
 
\subsection{Information Criteria and Model Selection}
\label{Criteria}

For the purpose of comparing a set of cosmological models regarding to their empirical 
predictions given the data, we use the Akaike Information Criterion (AIC)
\cite{Akaike1974}, the Bayesian Information Criterion (BIC) \cite{Schwarz1978}, and the
Deviance 
Information Criterion \cite{Spiegelhalter2002}.

The AIC criterion confronts the problem of model adequacy at the grounds of information 
theory. Specifically, it is an estimator of the Kullback-Leibler information with the 
property of asymptotically unbiasedness. Within the standard assumption of Gaussian 
errors, the AIC estimator is given by \cite{Ann2002,Ann2002b}
\begin{equation}
 \text{AIC}=-2\ln(\mathcal{L}_{\text{max}})+2k+
 \frac{2k(k+1)}{N_{\rm tot}-k-1}\,,
 \end{equation}
where $\mathcal{L}_{\text{max}}$ is the maximum likelihood of the data set(s) 
under consideration and $N_{\rm tot}$ is the total number of data points. Naturally, 
for large number of data points $N_{\rm tot}$, this expression
 reduces to $\text{AIC}\simeq -2\ln(\mathcal{L}_{\text{max}})+2k$, which 
corresponds to the ubiquitous form of the AIC criterion. Thus, it is preferable to use 
the 
modified AIC criterion in all cases \cite{Liddle:2007fy}.
        
The BIC criterion is an estimator of the Bayesian evidence (see e.g. 
\cite{Ann2002,Ann2002b,Liddle:2007fy} and references therein), and is given as
\begin{equation}
\text{BIC} = -2\ln(\mathcal{L}_{\text{max}})+k \,{\rm log}(N_{\text{tot}})\,.
\end{equation}
The DIC criterion is formulated using concepts from both Bayesian statistics and 
information theory \cite{Spiegelhalter2002} and is given as  \cite{Liddle:2007fy}
\begin{equation}
{\rm DIC} = D(\overline{\phi^\mu}) + 2C_{B},
\end{equation}
where $C_{B}$ is referred as Bayesian complexity. In particular, 
$C_{B} = \overline{D(\phi^\mu)} - D(\overline{\phi^\mu})$,
where the overline denotes the usual mean value.
Additionally, $D(\phi^\mu)$ is the Bayesian Deviation, which for a general class of 
distributions, 
that 
is the exponential family, it corresponds to $D(\phi^\mu) = 
-2\ln(\mathcal{L(\phi^\mu)})$.  This quantity is closely connected to the effective 
degrees of freedom  \cite{Spiegelhalter2002}, which  is the 
number of parameters that actually contribute to the fitting. To illustrate this, 
considering a model with a set of free parameters S, and a data set D, it is possible 
that we will be able to constrain only a subset of S. While AIC and BIC criteria will 
penalize the model using the total number of free parameters, DIC criterion will ``count" 
only the effective number of parameters in the context of D. Moreover, DIC utilizes the 
full log-likelihood sampling instead of just the maximum. In theory, employing only the 
likelihood value at the peak in our Bayesian framework could reduce the accuracy of the 
$\mathcal{L}_{max}$, as we calculate the mean value of the likelihood inside the 
$1\sigma$ 
area. However, by using ``long'' chains, we obtain $\mathcal{L}_{max}$ values with enough 
accuracy to use them to calculate AIC and BIC. An appealing feature of DIC is that, given 
the MCMC samples, its calculation is computationally cheap.

Given a set of rival models, our task is to rank the models at hand according to their 
fitting quality at the empirical data. We utilize the criteria presented previously, and 
more specifically the relative difference of the IC value for the given set
of models, $\Delta 
\text{IC}_{\text{model}}=\text{IC}_{\text{model}}-\text{IC}_{\text{min}}$,
where the $\text{IC}_{\text{min}}$ is the minimum $\text{IC}$ value in the set of 
competing models. 
We assign ``probability of correctness" to each model using the following rule
\cite{Ann2002,Ann2002b}:
\begin{equation}
\label{prob_per_model}
P \simeq \frac{e^{-\Delta \text{IC}_{i}}}{\sum_{i=1}^{n}e^{-\Delta 
\text{IC}_{i}} },
\end{equation}
where $i$ runs over the set of $n$ models under consideration.
In a direct analogy to the Bayes ratio \cite{Jeffreys}, the quantity
 $\Delta \text{IC}_{1}/\Delta \text{IC}_{2}$ could be thought as a measure of the 
relative strength of observational support between the two models. 
Further, in the context of the Jeffreys scale, as defined in \cite{Kass:1995loi}, 
the condition $\Delta\text{IC}\leq 2$, corresponds to statistical compatibility of the 
model at hand with the most favoured model by the data, while the condition 
$2<\Delta\text{IC}<6$ implies a middle tension between the two models, and the condition 
$\Delta\text{IC}\geq 10$ suggests a strong tension.

\section{Results}
\label{results}

In this section we confront $f(T)$ gravity, and in particular the three models 
$f_{1}$CDM, $f_{2}$CDM, $f_{3}$CDM presented in subsection \ref{fTmodels}, with the above 
observational datasets, following the aforementioned methods. 
The results for the 
parameters  are summarized in Table 
\ref{tab:Results1}. Additionally, in Figs. \ref{f_1_all}, 
\ref{f_2_all}, \ref{f_3_all} we present the corresponding  contour 
plots for each model respectively.

Comparing our $\textrm{SNIa}+\textrm{H(z)}+f\sigma_{8}$ results with the corresponding 
ones obtained  using $H(z)$ and 
Standard Candles data in \cite{Basilakos:2018arq}, we report $\sim 1 \sigma$ 
compatibility in all cases. However, in the present work we have obtained $\sim 40 \%$ 
smaller $b$ values, and $\sim 17 \%$ larger    matter-energy densities values,   while 
the error bars are about the same. A possible interpretation of the 
matter - energy density increment could be that it arises from the Pantheon SN Ia 
sample that we use here  instead of the  JLA one (see \cite{Scolnic:2017caz} for this 
effect on $\Lambda$CDM). Note that in our current work we do not use the 
$\gamma$-parametrization that was used in \cite{Nesseris:2013jea}, and thus although the 
$f(T)$ models analyzed in both works are the same, the statistical models are not.

Further, regarding our results using the  $SNia+H(z)+f\sigma_{8}+CMB_{\text{shift}}$ 
dataset, we observe $1\sigma$ compatibility of all acquired parameter values with the 
corresponding values obtained from late Universe data only. As for the model selection 
criteria, DIC criterion advocates on behalf $f_2$ model, while BIC and AIC criteria 
suggest the concordance model. Again, in contrast with other works (i.e 
\cite{Basilakos:2018arq}) $f_2$ and $f_3$ models are better than $f_1$.

\begin{table}[!]
\tabcolsep 4.0pt
\vspace{1mm}
\begin{tabular}{ccccccc} \hline \hline
Model & AIC & $\Delta$AIC & BIC &$\Delta$BIC & DIC & $\Delta$DIC
 \vspace{0.05cm}\\ \hline
%----------------------------------------
\hline
%  \vspace{0.25cm}
 \\
 \multicolumn{7}{c}{\emph{$H(z)$ + \text{SNIa} + \text{$f\sigma_{8}$}}}\\
  \vspace{0.05cm}\\ 
 $f_1$CDM &  77.658 & 2.184 & 89.631 & 4.482 & 75.572 & 0.609 \\  %\vspace{0.01cm}\\ 
$f_2$CDM &  79.689  & 4.214 &  91.663 & 6.513 & 75.185 & 0.222 \\ 
$f_3$CDM &  78.456  & 2.982 &  90.430  & 5.281 & 74.736 & 0 \\
$\Lambda$CDM & 75.474 & 0 & 85.149 & 0 & 74.963 & 0.227  \\ 
 \vspace{0.01cm}\\ 
  \multicolumn{7}{c}{\emph{$H(z)$ + \text{SNIa} + \text{$f\sigma_{8}$} + 
$\text{CMB}_{\text{shift}}$}}   \\
 \vspace{0.01cm}\\
 $f_1$CDM &  80.651  & 1.874 & 95.093 & 4.161 & 79.464 & 1.954 \\  %\vspace{0.01cm}\\ 
$f_2$CDM &   82.282  & 3.505 &  96.501 & 5.569 & 77.510 & 0 \\ 
$f_3$CDM &  81.944  & 3.167 &  96.163  & 5.231 & 77.539 & 0.029 \\
$\Lambda$CDM & 78.777 & 0 & 90.932 & 0 & 78.062 & 0.552  \\ 
\\ 
\hline\hline
\end{tabular}
\caption{The information criteria 
AIC, BIC and DIC for the examined cosmological models,
along with the corresponding differences
$\Delta\text{IC} \equiv \text{IC} - \text{IC}_{\text{min}}$.
\label{tab:Results2}}
\end{table}
 
We close this section by testing the statistical significance of our constraints. We 
implement the AIC, BIC and DIC criteria described in subsection \ref{Criteria}, and we 
present the results in Table \ref{tab:Results2}. A general conclusion is that the 
concordance $\Lambda$CDM 
paradigm seems to be favoured by the AIC and BIC criteria. However,  within the 
Deviance Information 
Criterion, the $f_3$CDM model seems to be favoured with a  very small difference 
between the competing models, and this is a novel result comparing to previous 
observational works, where $f_1$CDM model seemed to be the most favoured one 
\cite{Nesseris:2013jea,Basilakos:2018arq}. As we discussed in section 
\ref{DataMethodology}, in general the DIC criterion is more credible than the other two 
since it considers the \textit{effective} parameters number and moreover it takes into 
account the whole amount of information available from the sampling of the likelihood. 
Due to the small $\Delta\text{IC}$ differences ($\ll 2$) between the competing models, we 
are not in position to discriminate between them, and thus they can be considered as 
statistically equivalent.

 \section{Conclusions}
 \label{Conclusions}
 
 In this work we used observational data from Supernovae (SNIa) Pantheon sample, from 
direct measurements of the Hubble parameter that is cosmic chronometers (CC), from 
the 
Cosmic Microwave Background shift parameter $\text{CMB}_{\text{shift}}$,  and from 
redshift space distortions measurements ($f\sigma_8$), in order to constrain $f(T)$ 
gravity. Additionally, we did not follow the common $\gamma$ 
parameterization within the semi-analytical approximation of the growth rate, in order to 
avoid model dependent uncertainties. Up to our knowledge this is the first time that   
$f(T)$ gravity is analyzed within a Bayesian framework, and with background and 
perturbation behaviour considered jointly.

We considered three $f(T)$ models, which are viable  since they pass the basic 
observational tests, and we quantified their deviation from $\Lambda$CDM cosmology
through a sole parameter. 
Our analysis revealed that these $f(T)$ models  are able to describe 
adequately the $f\sigma_8$ data. Furthermore, by applying AIC and BIC criteria  we 
deduced that $\Lambda$CDM cosmology is 
still favoured by the  CC+SNIa+$f\sigma_{8}$+ $\text{CMB}_{\text{shift}}$       
joint analysis. 
The extracted parameter values are in good agreement with previous 
observational analyses which used only background data  \cite{Basilakos:2018arq},
however an interesting finding is that while the previous works favoured $f_1$CDM model, 
the present investigation seems to favour  $f_3$CDM one.

Finally, applying the more efficient  DIC criterion we saw that the smallness of  
$\Delta\text{IC}$  suggest statistical equivalence 
between $f_2$CDM, $f_3$CDM and the concordance $\Lambda$CDM cosmology. This could offer a 
motivation for using these two models for developing a new,   more competitive $f(T)$ 
 scenario. In summary,  $f(T)$ modified gravity is a good candidate for the description 
of nature and deserves further investigation.

\section*{Acknowledgments}
 This article is based upon work from COST Action ``Cosmology and Astrophysics Network
for Theoretical Advances and Training Actions'', supported by COST (European Cooperation
in Science and Technology).

\end{document}